\documentclass[a4paper,11pt]{article}
\usepackage{jcappub,aas_macros} 
\usepackage{lineno}
\usepackage{stmaryrd}

\pdfoutput=1 

\usepackage{mathtools}

\usepackage[normalem]{ulem}

\newcommand{\edit}[1]{{#1}}

\title{It's All {\tt Ok}: Curvature in Light of BAO from DESI DR2}

\author[a]{Shi-Fan Chen}
\author[a]{Matias Zaldarriaga}
\affiliation[a]{Institute for Advanced Study, 1 Einstein Drive, Princeton, NJ 08540, USA}

\emailAdd{sfschen@ias.edu}

\abstract{Recent measurements of baryon acoustic oscillations (BAO) from the Dark Energy Spectroscopic Instrument (DESI) show hints of tension with data from the cosmic microwave background (CMB) when interpreted within the standard model of cosmology. In this short note we discuss the consequences of one solution to this tension, a small but negative spatial curvature with $R_k = 21 H_0^{-1}$, which DESI measures at $2\sigma$ \edit{when combined with CMB data}. We describe the physical role of curvature in cosmological distance measures tied to recombination, i.e. the CMB and BAO, and the relation to neutrino mass constraints which are relaxed to $\sum m_\nu < 0.10$ eV \edit{at $95\%$ confidence} when curvature is allowed to deviate from zero. A robust detection of negative curvature would have significant implications for inflationary models: improved BAO measurements, particularly from future high-redshift spectroscopic surveys, will be able to distinguish curvature from other solutions to the DESI-CMB tension like phantom dark energy at high significance.}

\begin{document}
\maketitle
\flushbottom

\section{Introduction}
\label{sec:intro}

In the past two decades, the combination of precision measurements of the cosmic microwave background (CMB) \cite{WMAP,Planck18,ACTDR6} and large-scale structure---particularly through the measurement of baryon acoustic oscillations (BAO)\cite{eBOSS}---have painted a picture of a universe surprisingly well described by a simple standard model of cosmology with a flat FLRW metric governed in the late universe by dark energy ($\Lambda$), cold dark matter (CDM) and baryons. Recently, the Dark Energy Spectroscopic Instrument (DESI) released the tightest BAO measurements to date using galaxies and the Lyman-$\alpha$ forest in their second data release (DR2) \cite{BAODR2,DR2LyA}. These new data show surprising signs of tension with current CMB data, manifesting for example as a great\edit{er} than $2\sigma$ tension in the current matter fraction $\Omega_M$ when fit to the flat $\Lambda$CDM model \edit{with the minimum allowed neutrino mass}. 

Indeed, the DESI collaboration showed that these tensions can be resolved, and the goodness of fit improved, if the cosmological constant in the standard model is replaced by a dynamical dark energy (DDE) component, which they model using an empirical equation of state $w(a) = w_0 + w_a(1 - a)$ linear in the scale factor $a$. These data prefer this $w_0 w_a$CDM extension of the standard model of cosmology $\Lambda$CDM by more than $3\sigma$ when considering only data from cosmic perturbations (CMB and BAO), with even higher significance attained when combining with Type IA supernovae. Intriguingly, the DESI CMB constraints prefer DDE where the equation of state $w < -1$ at redshifts $z > 0.5$, forcing dark energy to enter a (seemingly) unphysical \textit{phantom} regime where its energy density increases with \edit{time}, violating the null energy condition \cite{Lodha25}.

Beyond DDE, within $\Lambda$CDM, DESI + CMB also places stringent constraints on neutrino mass. Taken at face value, these constraints show a preference for the sum of the neutrino masses $M_\nu = \sum m_\nu$ to be below that required from neutrino oscillation experiments $0.06 \text{eV}$ at the close to $95\%$ level. This preference is enhanced if the ``effective'' neutrino mass is allowed to take on negative values \cite{Elbers25}. However, since DESI's contribution to the neutrino mass constraint is geometric, relaxing the cosmological expansion history in the standard model by, for example, allowing for DDE significantly relaxes these constraints.

\begin{figure}
    \centering
    \includegraphics[width=0.8\linewidth]{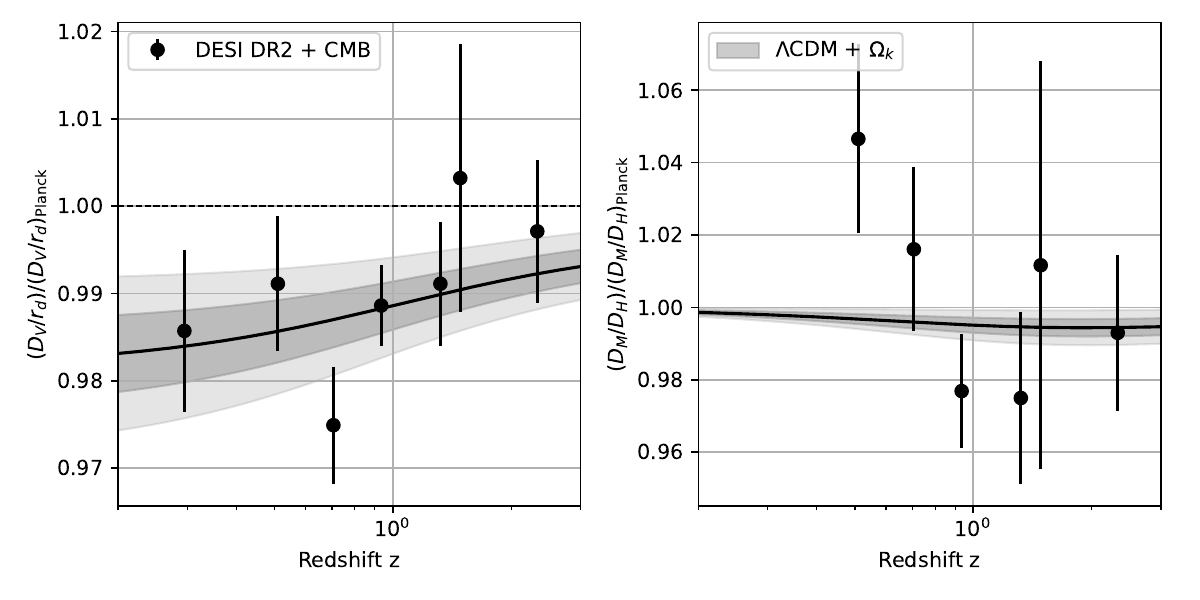}
    \caption{Predictions for $D_V/r_d$ and $D_M/D_H$ in a curved $\Lambda$CDM universe, normalized to those predicted in the best-fit Planck $\Lambda$CDM cosmology, based on the joint fit of DESI DR2 BAO and CMB data. The DESI DR2 measurements are shown in black, while gray bands show $1$ and $2\sigma$ regions based on these data. Note that predictions for each model in these regions are smooth, i.e. the gray bands should be interpreted to be correlated and smoothly varying across redshifts $z$.}
    \label{fig:ppd}
\end{figure}

The apparent preference for unphysical behavior from either dark energy or neutrinos suggests re-evaluating common assumptions about the standard cosmological model. Our purpose in this work is to explore the implications for one physical extension of the standard model, i.e. the possibility of nonzero spatial curvature ($\Omega_k$). Indeed, the DESI collaboration reported constraints on a curved $\Lambda$CDM + $\Omega_k$, finding a $2\sigma$ preference for a negatively curved universe with $\Omega_k = 0.0023 \pm 0.0011$ \edit{when combining their measurements with CMB data} \cite{BAODR2}.\footnote{\edit{The DESI collaboration also reports constraints on the spatial curvature when allowing for dynamical dark energy, finding that the combination of DESI, CMB and supernovae data prefer a spatially flat universe within $1\sigma$ when dark energy in the phantom regime is allowed \cite{BAODR2}.}} The implications of this preference for the local distance scale, compared to the flat $\Lambda$CDM model preferred by CMB data from the Planck satellite \cite{Planck18}, are shown in Figure~\ref{fig:ppd}. While statistically somewhat weaker than the $3\sigma$ preference for DDE, this suggestion of nonzero curvature is nonetheless interesting because it is a physically well-understood solution to the BAO-CMB tension compared to phantom dark energy or negative neutrino masses. Indeed, the small negative value of the spatial curvature preferred by BAO+CMB data has direct implications for for inflationary physics (see e.g.~\cite{Planck18_Inflation} for a review): a detection of positive curvature would be difficult to reconcile with naive expectations of inflation, while a negative but nonzero ($> 10^{-4}$) one as detected here would rule out slow-roll eternal inflation while remaining consistent with open inflation scenarios, e.g. involving bubble nucleation \cite{Kleban12}.

The rest of the paper is structured as follows. We give an overview of spatial curvature in Friedmann–Lemaître–Robertson–Walker (FLRW) spacetimes, and how it can resolve the tension between the CMB and BAO at high and low redshifts, in Section~\ref{sec:curvature}. We then describe the implications of curvature at this level for the cosmological determination of the neutrino mass in Section~\ref{sec:neutrinos}, and for detection by future galaxy surveys in Section~\ref{sec:detectability}, before concluding in Section~\ref{sec:discussion}. Throughout, we will use BAO data from DESI DR2 and, for the CMB, a combination of the low $\ell$ data from the PR3 of the Planck satellite \cite{Planck18Like}, high $\ell$ \texttt{Camspec} from PR4 \cite{NPIPE_Camspec}, and a combination of Planck and Atacama Cosmology Telescope (ACT) CMB lensing data using the publicly available likelihood by the ACT collaboration \cite{PR4_Lensing,ACT_Lensing_a,ACT_Lensing_b}. Our constraints are run using \texttt{Cobaya} \cite{Cobaya} using theory predictions from \texttt{CAMB} \cite{CAMB} and plotted using \texttt{GetDist} \cite{GetDist}.

\section{The Role of Curvature}
\label{sec:curvature}

\begin{figure}
    \centering
    \includegraphics[width=0.8\linewidth]{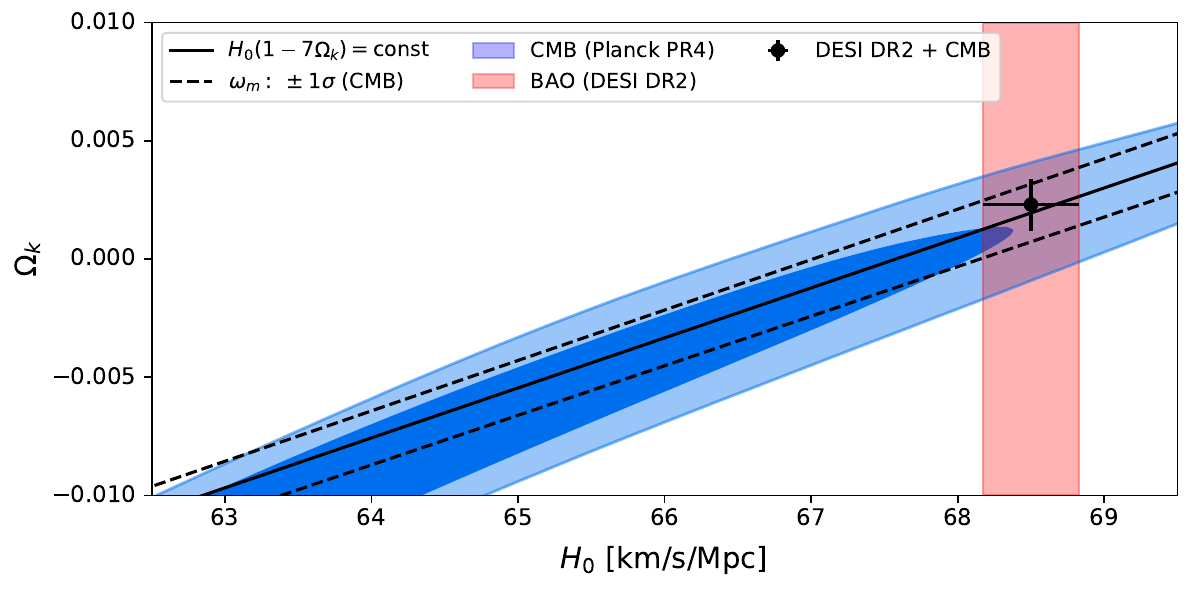}
    \caption{Degeneracy between $\Omega_k$ and $H_0$. The constraint from CMB temperature and polarization spectra, shown here from Planck PR4 chains \cite{NPIPE_Camspec}, follows the $H_0 (1 - 7 \Omega_k)$ degeneracy at small values of the curvature, with a width set by uncertainty in $\omega_m$ from the primary CMB \cite{Lemos23}. With this same $\omega_m$ constraint, DESI BAO places a strong bound on $H_0$, leading to an $\Omega_k$ constraint mostly dictated by CMB uncertainties. }
    \label{fig:degeneracy}
\end{figure}

Baryon acoustic oscillations (BAO) probe the expansion history in the late ($z \approx 0-2$) universe by providing a standard ruler with length given by the sound horizon at the end of the baryon drag epcoh $z_d \approx 1100$
\begin{equation}
    r_d = \int_{z_d}^\infty \frac{dz\ c_s(z)}{H(z)},
\end{equation}
where $c_s(z)$ is the sound speed and $H(z)$ is the Hubble parameter at redshift $z$, in the form of a peak in the 2-point function of galaxies. Converting into the observed angular and redshift coordinates of galaxy surveys, BAO then directly constrain the dimensionless quantities $D_M(z) / r_d$ and $H(z) r_d$, where $D_M$ and $H(z)$ are the transverse comoving distance and Hubble parameter, at the effective redshift $z$ of the galaxies. These measurements are closely related to the measurement of the angular acoustic scale of the CMB, which probe\edit{s} the ratio of the ratio of the transverse comoving distance and sound horizon at \edit{d}ecoupling $\theta_\ast = D_{A,\ast}/r_\ast$. This angular scale is the best-measured quantity from CMB spectra, constrained for example by Planck to three parts per ten thousand by the Planck satellite \cite{Planck18}. 

Within the standard model of cosmology, CMB data alone are sufficient to give reasonably tight predictions for BAO measurements at lower redshifts. This is because the primary CMB by itself constrains the physical baryon and cold dark matter densities $\omega_b, \omega_{cb} = \Omega_b h^2, (\Omega_c +\Omega_b) h^2$, which together determine $r_d$, and the remaining degree of freedom, the present-day Hubble constant $H_0 = 100 \ h\ \text{km/s/Mpc}$, is fixed by the constraint on the distance-to-last-scattering $D_M(z_\ast)$ set by $\theta_\ast$, since the dark energy fraction $\Omega_\Lambda$ is fixed given any set of these three parameters. In this way of thinking, the roughly $1.5\%$ shorter distances compared to Planck $\Lambda$CDM, as measured by DESI BAO, can be roughly thought of as a small $1.5\%$ Hubble tension at $z \lesssim 1.5$. Indeed, the DESI collaboration show that primary CMB constraints on $\omega_b, \omega_{cb}$ and $\theta_\ast$ alone, as presented in ref.~\cite{Lemos23}, are sufficient to reproduce most of the preference for DDE when combined with DESI BAO \cite{BAODR2}. The former two densities are constrained by the primary CMB from Planck to roughly $0.7\%$ and $0.84\%$, respectively, meaning that absent changes to recombination physics the cosmological uncertainty on $r_d$ is at the less than $0.2\%$ level. \edit{Freeing up spatial curvature as we do here does not affect CMB-era physics and leads to essentially no change in the miniscule uncertainty on $r_d$.}

Spatial curvatures relaxes the tight relations set by CMB constraints on $\omega_b$, $\omega_c$ and $\theta_\ast$ by introducing a small nonlinearity into the relations between cosmological distances. Since the value of $\Omega_k$ required to explain the tension between DESI and the CMB is so small compared to $\Omega_{m,\Lambda}$, it plays only a very small role in determining $H^2(z) \sim \rho(z)$ at any epoch. A given value of $\Omega_k$ corresponds to a radius of curvature in comoving coordinates of
\begin{equation}
    R_k = \frac{1}{\sqrt{|\Omega_k|}H_0} \approx 21 H_0^{-1} \left( \frac{|\Omega_k|}{0.0023} \right)^{-\frac12}.
\end{equation}
In comparison, the comoving distances to redshifts $z=0.5, 1, 1100$ are roughly $0.4 H_0^{-1}, 0.8 H_0^{-1}$ and $3.1 H_0^{-1}$, respectively. The fractional difference between the comoving distance $\chi$ and transverse comoving distance $D_M = R_k \sinh(\chi/R_k)$ is given by $\chi/D_M \approx 1 + \frac16 R_k^{-2} \chi^2$ to leading order and negligible for the redshifts probed by BAO for the values of curvature we consider. The correction at last scattering is roughly equal to $\frac53 \Omega_k$ for $\Omega_m \approx 0.3$, or \edit{$0.35\%$} at $\Omega_k = 0.0023$. All else being equal\footnote{Here, for a given $\Omega_k$ we keep the $H_0$ and $\Omega_{m} h^2$ constant while adjusting $\Omega_\Lambda$.}, the comoving distance to last scattering $\chi_\ast$ is changed by a smaller relative amount, roughly $- \Omega_k / 4$, leading to a total change \edit{$\Delta D_{M,\ast}/D_{M,\ast} \approx 1.4\ \Omega_k. $}

We can thus think of curvature, at the values allowed by current data, as only modifying the distance to last scattering in $\theta_\ast$. This allows us to measure $\Omega_k$: In flat $\Lambda$CDM, the angular acoustic scale roughly constrains the combination $\Omega_m h^3$, with $\theta_\ast$ scaling roughly as $\omega_m^{0.14} h^{0.2}$ \cite{Hu01,Percival02}, where $\omega_m = \Omega_m h^2$. Accounting for changes in the transverse distance due to curvature adds a factor of roughly $(1 - 1.4 \Omega_k)$ to $\theta_\ast$, and modifies the best-fit combination given our choice of data
\begin{equation}
    \Omega_m h^3 (1 - 7 \Omega_k) = 0.09603 \pm 0.00026.
\end{equation}
However, the primary CMB independently constrains $\omega_m$ to sub-percent accuracy, allowing us to constrain the combination $H_0 (1 - 7 \Omega_k) $ to roughly $\sigma(\omega_m)/\omega_m \lesssim 1\%$.

At low redshifts, the effects of curvature are negligible, so BAO constrain $H_0$ through $H_0 r_d$---assuming that variations in $r_d$ are subdominant---and $\Omega_m$ almost identically to the flat $\Lambda$CDM case. Indeed the DESI data best constrain a combination close to $\Omega_m^{0.2} H_0 = \omega_m^{0.2} H_0^{0.6}$, at roughly the $0.3\%$ level. Assuming again a CMB prior on $\omega_m$, this is equivalent to a roughly $0.5\%$ constraint on $H_0$, allowing CMB + BAO to break the $H_0-\Omega_k$ degeneracy and constrain BAO. The factor of $7$ in the degeneracy gives us a curvature measurement of about $\sigma(\Omega_k) = 0.0012$. The slightly higher $H_0$ measured by BAO\edit{, calibrated using information from the CMB on the physical density of baryons and dark matter, relative to that implied by the CMB directly in $\Lambda$CDM} is thus efficiently translated into a curvature of $\Delta H_0 / (7 H_0) \approx 0.002$. This degeneracy is illustrated in terms of the $H_0-\Omega_k$ posterior in Figure~\ref{fig:degeneracy}. Interestingly, the BAO constraint on $H_0$ given a CMB prior on the \edit{physical} matter density is slightly tighter than the constraint on $H_0 (1 - 7 \Omega_k)$ from $\theta_\ast$ given the same prior, meaning the error is mainly set by the CMB uncertainty in $\omega_m$. \edit{This reflects that only the large distances close to the CMB are meaningfully affected by small values of curvature, such that the $\theta_\ast$ constraint probes a degeneracy direction strongly dependent on $\Omega_k$ while low-redshift BAO probe parameter combinations that are essentially independent. Merely tightening the error bar around DESI BAO measurements will thus only somewhat improve curvature constraints.}

\begin{figure}
    \centering
    \includegraphics[width=0.8\linewidth]{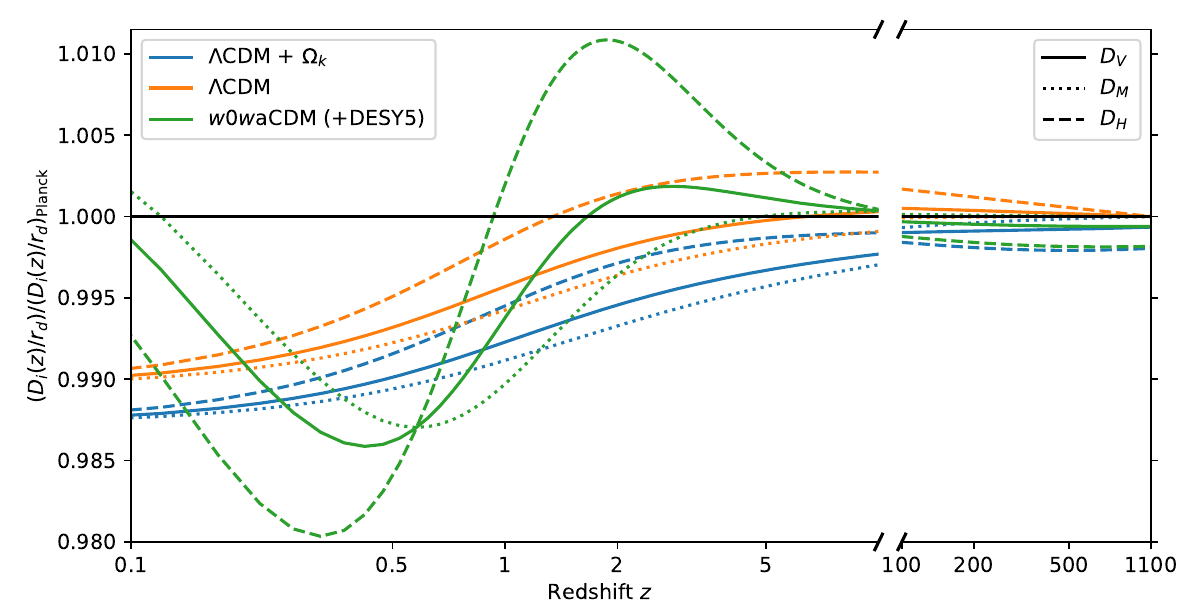}
    \caption{Distances $D_H$, $D_M$ and $D_V$ fit to three ($\Lambda$CDM, curved $\Lambda$CDM and $w_0 w_a$CDM) cosmological models given DESI and CMB data, normalized to their values in the Planck best-fit $\Lambda$CDM cosmology. The right extension shows these distances at high redshift, where $D_M/r_d$ in all models is forced to converge by the CMB constraint on $\theta_\ast$. }
    \label{fig:distances}
\end{figure}

Figure~\ref{fig:distances} shows $D_{V,M,H}$ in different cosmological models fit to DESI BAO and the CMB at the low and high redshifts measured by the former and latter, respectively. In each of these models the best-fit parameters conspire to decrease the measured BAO scale by about $1.5\%$ between the $z \approx 0.5 - 1.5$, reflecting the low points in Figure~\ref{fig:ppd}. However, one notable difference is that the best fit curved $\Lambda$CDM model exhibits a fairly smooth dependence at low redshifts, amounting to a rescaling relative to the best-fit flat model, while the \edit{best-constrained} DDE $w_0 w_a$CDM model \edit{reported by DESI, including supernovae data from DES Y5,} induces relatively large changes in the energy density and distances at low redshifts to fit the data. \edit{These changes allow the $w_0 w_a$CDM model to better fit the ratio between the line-of-sight ($D_H$) and transverse ($D_M$) distances, as can also be seein in the right panel of Figure~\ref{fig:specs5}, at the expense of allowing dark energy to cross into the phantom regime.}

\section{Implications for Neutrino Mass}
\label{sec:neutrinos}

In the standard model, the combination of CMB and BAO data allows us to geometrically measure the neutrino mass using the same degeneracy breaking as discussed above in Section~\ref{sec:curvature}. We point the interested reader to ref.~\cite{Weiner24}, as well as the DESI collaboration's supporting paper on neutrino mass constraints \cite{Elbers25}, for a more detailed discussion; briefly: since neutrinos are still relativistic at recombination, the CMB power  spectrum shape constraint on $\omega_m$ is really a constraint on $\omega_{cb} = (1 - f_\nu) \omega_m = (\Omega_c + \Omega_b) h^2$, where $f_\nu$ is the neutrino mass fraction, in the presence of massive neutrinos. As in the case of $\Omega_k$, geometric data in the CMB and BAO then allows us to distinguish between $\omega_{cb}$ and $\omega_m$ \edit{since the distance to last scattering depends on the matter density including neutrinos}, giving us a constraint on $M_{\nu} \approx (\omega_m - \omega_{cb})\ 93\ \text{eV}$. For example, in the case of the angular acoustic scale, the addition of massive neutrinos decreases the distance to last scattering \edit{$D_{M,\ast} \propto \Omega_m^{-0.1}/\sqrt{\omega_m}$} by a factor of roughly $(1 + 0.4 f_\nu)$ compared to a neutrino-less universe with the same $H_0$ and baryon and dark matter densities \cite{Percival02,Weiner24}, leading to $\theta_\ast \sim \omega_{cb}^{0.14} h^{0.2} (1 + 0.4 f_\nu) $ and a best-constrained combination $\Omega_m h^3 (1 + f_\nu)$.\footnote{Empirically, we find for our data including CMB lensing that the best-constrained combination is modified to $\Omega_m h^3 (1 + 0.65 f_\nu) = 0.09628 \pm 0.00026$.} This additional degree of freedom implies that the primary CMB can no longer break all the degeneracies by itself, but the measurement of the BAO scale at low redshift can as in the case of curvature.

\begin{figure}
    \centering
    \includegraphics[width=0.8\linewidth]{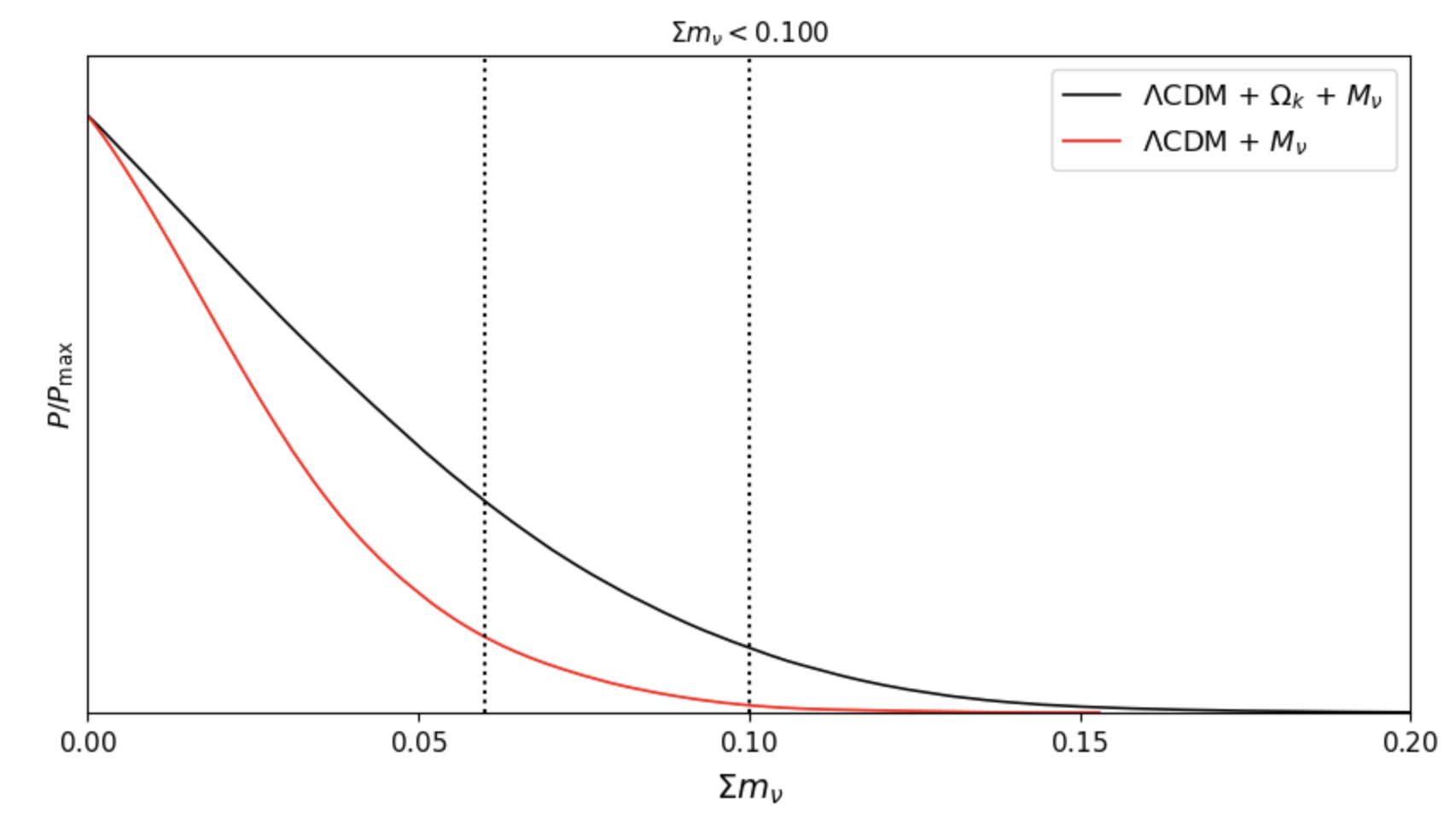}
    \caption{The 1D posterior for the sum of the neutrino masses assuming a positivity prior in a $\Lambda$CDM (red) and curved $\Lambda$CDM (black) universe. Neutrino mass constraints are significantly relaxed in the latter case, with both the normal and inverted hiearchies (black dotted lines) allowed at $95\%$ confidence.}
    \label{fig:mnu_constraints}
\end{figure}

Allowing for nonzero spatial curvature relaxes this constraint because the curvature is determined through breaking the same degeneracy. Roughly speaking, the degeneracy direction is determined by their respective modifications to $\theta_\ast$: keeping $\theta_\ast$ constant requires $1.4 \Omega_k = 0.4 f_\nu$, i.e. the degeneracy has a steep slope of $\Delta f_\nu / \Delta \Omega_k = 3.5$. 

\begin{figure}
    \centering
    \includegraphics[width=0.8\linewidth]{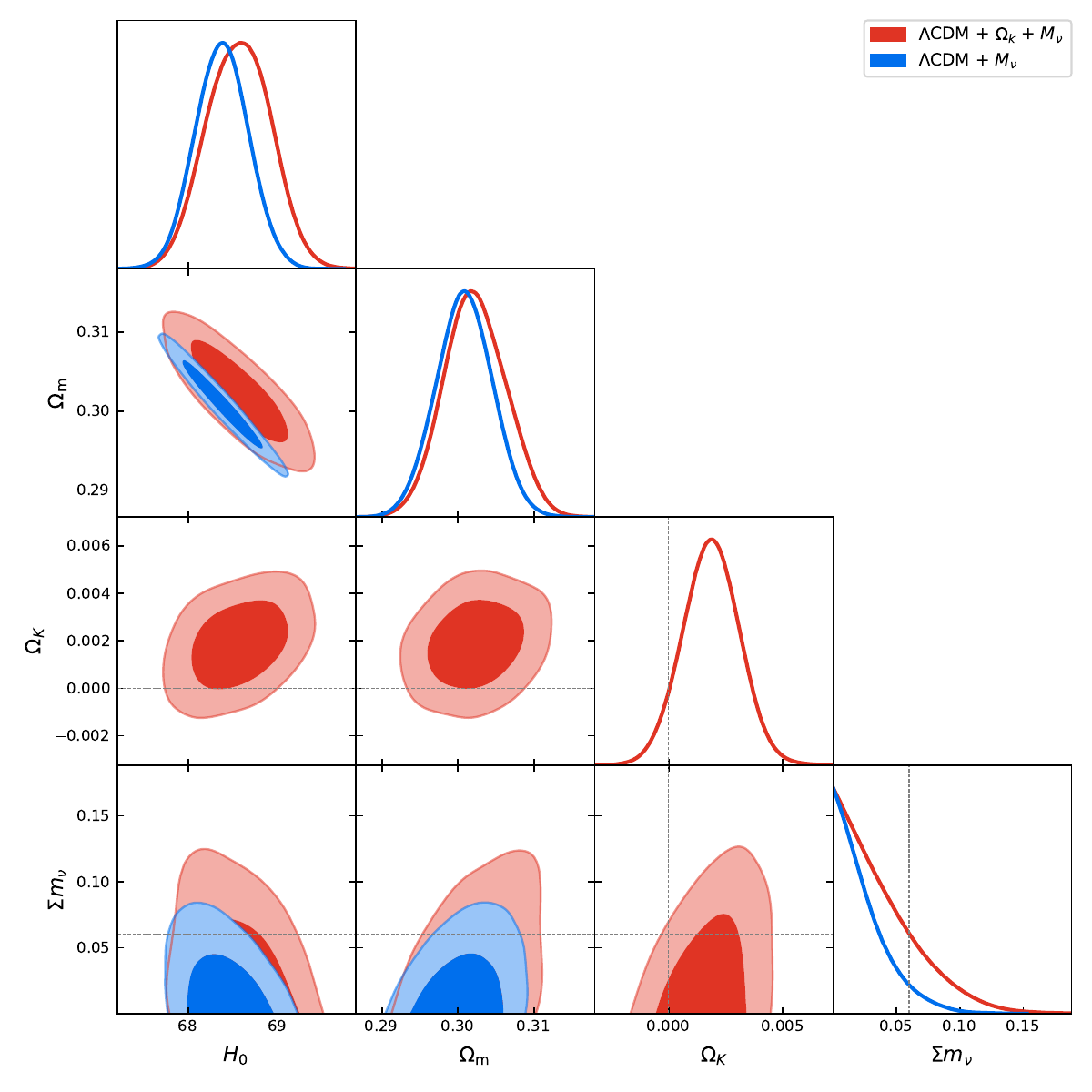}
    \caption{Same as Fig.~\ref{fig:mnu_constraints}, but for two-dimensional posteriors with other cosmological parameters.}
    \label{fig:mnu_contours}
\end{figure}

Figures~\ref{fig:mnu_constraints} and \ref{fig:mnu_contours} show the 1- and 2-D posteriors of the curved $\Lambda$CDM model with a free neutrino mass. We adopt a positivity prior on the neutrino mass for brevity. Allowing for spatial curvature significantly relaxes the constraints on $M_\nu$, such that both the normal and inverted hiearchies fall within the $95\%$ confidence interval with this choice of prior\edit{. However, it is important to note that the degeneracy between $\Omega_k$ and $M_\nu$ is not exact since BAO also constrain the shape of the expansion history and additional information in the full CMB, most importantly through CMB lensing, probe structure growth in the late universe. Indeed, while curvature constraints from CMB lensing combined with the primary CMB are very subdominant to the CMB and BAO combination \cite{Planck18,ACT_Lensing_a}, CMB lensing and BAO can constrain $M_\nu$ rather competitively with the geometric constraint discussed above using the distance to last scattering \cite{Weiner24}. The constraint on $\Omega_k$ from our full set of CMB and BAO data is } essentially unaffected by the inclusion of $M_\nu$ as a free parameter, owing to the steep slope of the $\Omega_k-f_\nu$ degeneracy and the preference of the CMB lensing + BAO for very low values of the neutrino mass even in the absence of geometric information from $\theta_\ast$ (see e.g. Fig.~8 of ref.~\cite{Weiner24} and Fig.~3 of ref.~\cite{Lynch25}), as can also be seen in the bottom right contour in Figure~\ref{fig:mnu_contours}.

\section{Detectability in Future Surveys}
\label{sec:detectability}

\begin{figure}[h]
    \centering
    \includegraphics[width=0.8\linewidth]{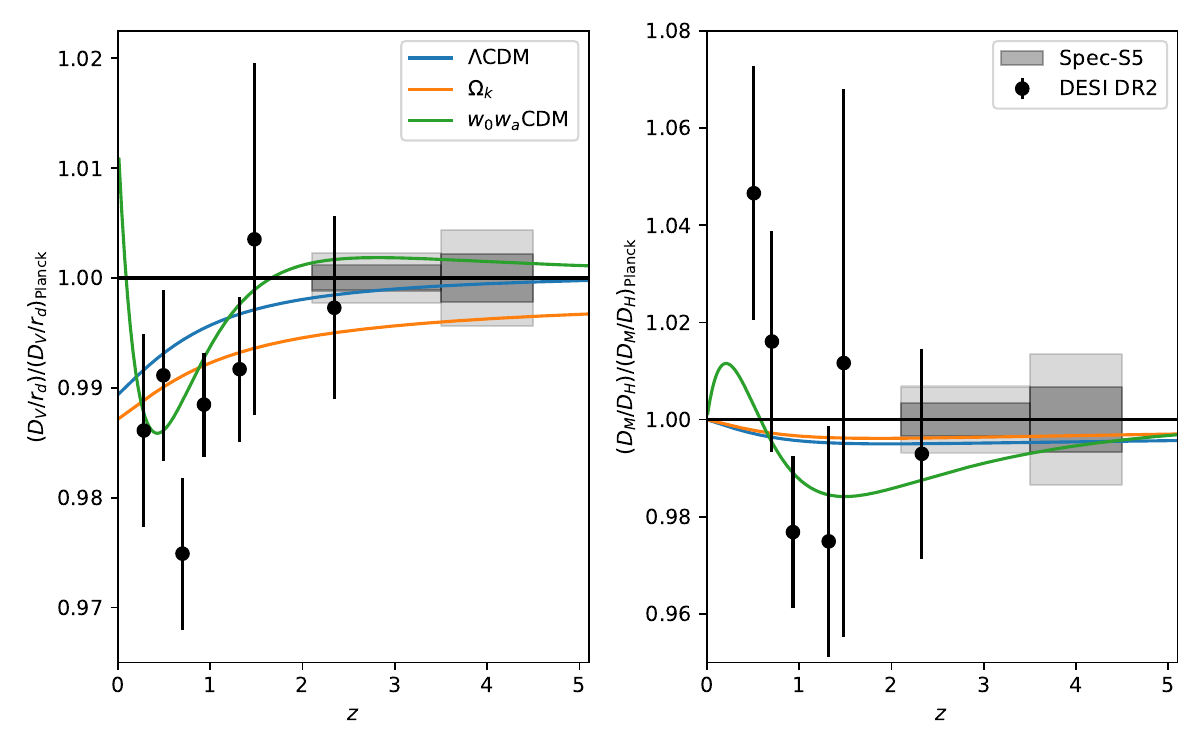}
    \caption{Detectability of different cosmological models in light of DESI DR2 (black points) and the planned Spec-S5 experiment (gray $1-2\sigma$ bands). At the redshifts probed by Spec-S5, a curved model preferred by the DESI+CMB data predicts shorter distances at the roughly $0.5\%$ level compared to $\Lambda$CDM or DDE, which Spec-S5 will be able to constrain at high significance. }
    \label{fig:specs5}
\end{figure}

As shown in Figure~\ref{fig:distances}, one of the distinguishing features of curvature as compared to other models that resolve the tension between DESI BAO and CMB data is that cosmological distances, and in particular the transverse comoving distance, can stay low relative to what is required to preserve $\theta_\ast$ until relatively high redshifts owing to the nonlinear relation between distance measures in cosmologies with spatial curvature. In order to investigate whether this signal can be meaningfully utilized in future cosmology experiments, let us consider the expected constraints from a proposed Stage-V spectroscopic instrument (Spec-S5) \cite{SpecS5}. We estimate these constraints using a Fisher matrix calculation based on two galaxy samples in redshift bins $z \in [2.1, 3.5],\ [3.5, 4.5]$ over $11,000$ square degrees.\footnote{We thank Martin White for providing the number densities and linear biases for these samples in ref.~\cite{SpecS5}.} Since nonlinear effects on the BAO are expected to be small at these redshifts, we adopt a simple model using linear theory, marginalizing over the linear bias and growth rate as well as broadband quintic polynomials in wavenumber $k$, for our forecasts, and assume $k_{\rm max} = 0.3\ h\text{Mpc}^{-1}$, though we find our forecasts to be relatively insensitive to these choices.

Figure~\ref{fig:specs5} shows the forecasted Spec-S5 constraints compared to the predictions of different cosmological models. Spec-S5 is expected to constrain the isotropic BAO scale $D_V/r_d$ over its combined redshift range at better than $0.1\%$: notably, while both the DESI + CMB best-fit $\Lambda$CDM and $w_0 w_a$CDM models would fit inside a $95\%$ confidence interval in both redshift bins, the curved $\Lambda$CDM curve would be detected at greater than $5\sigma$ using the isotropic BAO. This points to the significant promise of high-redshift spectroscopic observations to pin down the curvature of the universe while distinguishing it from other forms of beyond-$\Lambda$CDM physics. We note also that the constraints on the Alcock-Paczynski distortion $D_M/D_H$, while subdominant, can in addition measure the best-fit $w_0 w_a$CDM model at more than $2\sigma$.

\section{Conclusions}
\label{sec:discussion}

Cosmological observations point to a universe close to $\Lambda$CDM that is exceedingly flat. However, recent BAO data published by the DESI collaboration \cite{BAODR2}, when combined with CMB data, point to a $2\sigma$ hint of nonzero, negative spatial curvature, with a comoving curvature radius of about $21 H_0^{-1}$. A detection of negative spatial curvature would have significant implications for e.g. inflation, which cannot generically occur given positive curvature, and rule out scenarios such as slow-roll eternal  inflation.

In this note we have explored the consequences of such a small negative value of the curvature, or positive $\Omega_k$, for the CMB and large-scale structure. Freeing $\Omega_k$ at this level in cosmological analyses alleviates tension between the CMB and DESI BAO measurements by keeping the constraints from low-redshift BAO measurements and CMB power spectrum shape information essentially unchanged from flat $\Lambda$CDM while adding a steep dependence to the observe\edit{d} angular acoustic scale $\theta_\ast$. \edit{Freeing up curvature thus allows for shorter distances at lower redshifts, corresponding to a roughly $2\%$ higher $H_0$ compared to the value inferred from the CMB in flat $\Lambda$CDM, similar to that preferred in flat $\Lambda$CDM by DESI BAO combined with big-bang nucleosynthesis (BBN) constraints---though still notably lower than that measured using the local distance ladder \cite{BAODR2}.} This effect also significantly widens constraints on the sum of the neutrino masses $M_\nu$, whose geometric measurement from these data rely on the same set of physical quantities, such that even a present day $\Omega_k$ at two tenths of a percent are sufficient to widen $M_\nu$ constraints from modestly disfavoring the normal hierarchy to accommodating both the normal and inverted hierarchies within $95\%$ confidence. Curvature at these small levels can be detected at high significance by high-redshift spectroscopic surveys such as the proposed Spec-S5 experiment \cite{SpecS5} and distinguished from other scenarios such as dynamical dark energy, pointing to the significant promise of such surveys in unveiling beyond-$\Lambda$CDM physics.

Since the role of curvature in fits to BAO + CMB data is essentially to allow modifications to the distance to last scattering $D_{M,\ast}$ while keeping other cosmological parameters fixed, it is closely (inversely) related to alternative models that address tensions by changing the physics at recombination by changing the last-scattering sound horizon $r_\ast$ or redshift $z_\ast$. Ref.~\cite{Hamidreza25} recently explored this possibility in the context of DESI DR2 for a phenomenological model modifying recombination, and many such models were constrained using CMB power spectra from ACT DR6 \cite{ACT_extended}. Similarly to curvature, which will be probed by future high-redshift spectroscopic data, these modifications of recombination will be better constrained by future CMB data, including the Simons Observatory \cite{SO} and CMB-S4 \cite{CMBS4}. In the absence of new pre-recombination physics, these new observatories will significantly improve constraints on $\omega_b, \omega_{cb}$ which are currently a main limiting factor in constraining $\Omega_k$. Together with expected improvements in measurements of the BAO scale (e.g. $H_0 r_d$), these future experiments will help home in on present tensions between the two types of data. While we have kept with the data used in the DESI DR2 analyses in ref.~\cite{BAODR2}, it would be interesting to explore these avenues in light of the recent ACT DR6 CMB power spectra \cite{ACTDR6}---as done for other models in ref.~\cite{DESIACT}---for this same reason.

Finally, while we have focused on the role of spatial curvature in cosmic structure data, it is worth noting in closing a few implications for constraints on DDE using additional probes, in particular Type Ia supernovae \cite{DESY5,Union3,Pantheon}. The DESI collaboration combines these data, particularly those from ref.~\cite{DESY5}, with BAO and CMB data to yield a significant detection of DDE, and in particular DDE in the phantom regime, assuming a spatially flat universe. Recently, refs.~\cite{Bhattacharya24,Akrami25,Dinda25} conducted analyses of physical models of DDE with a free spatial curvature, finding that the data can be equally well fit by non-phantom dark energy in that scenario. This is because, in a cosmology with nonzero spatial curvature, the suppression of the distance scale at low redshifts compared to the CMB can be achieved via the nonlinear relation between different distances on large scales, rather than an unexpected increase in the late-universe energy density requiring dark energy to increase with time. Future observations, e.g. at high redshifts from Spec-S5, will be critical in differentiating between extensions of the standard model such as curvature or dynamical dark energy.

\section*{Acknowledgements}

We thank Colin Hill, Nickolas Kokron, Gabriela Sato-Polito and Giovanni Maria Tomaselli for useful discussions. We in particular thank Noah Sailer, Gerrit Farren, and Mat Madhavacheril for their assistance with cosmological likelihoods in \texttt{Cobaya}, and Martin White for providing the forecast properties of Spec-S5 galaxies.

SC and MZ acknowledge support from the National Science Foundation at the IAS through NSF-BSF 2207583. MZ is also supported by NSF 2209991
and the Nelson Center for Collaborative Research.

\bibliography{biblio}
\bibliographystyle{jhep}

\end{document}